\pdfoutput=1
\documentclass[twocolumn,a4paper,11pt,oneside]{article}
\usepackage[dvipsnames]{xcolor}
\usepackage{amsmath}
\usepackage[a4paper, margin=0.7in]{geometry}
\usepackage{amsfonts}
\usepackage{multirow}
\usepackage{accents}
\usepackage[switch,columnwise]{lineno}
\usepackage{hyperref}
\usepackage[english]{babel}
\usepackage[utf8]{inputenc}
\usepackage{url}
\usepackage[pdftex]{graphicx}
\usepackage{abstract}
\usepackage{mathtools}
\usepackage{float}
\usepackage{footnote}
\usepackage{etoolbox}
\usepackage{textcomp}
\newcommand{\authorblock}[1]{\begin{tabular}{@{}c@{}}#1\end{tabular}}

\usepackage[labelfont=bf,small,labelsep=space]{caption}
\newcommand{\ie}{\textit{i}.\textit{e}. }
\def\TReg{\textsuperscript{\textregistered}}

\newcommand{\argmax}{\mathop{\mathrm{argmax}}}
\newlength{\dhatheight}
\newcommand{\doublehat}[1]{%
    \settoheight{\dhatheight}{\ensuremath{\hat{#1}}}%
    \addtolength{\dhatheight}{-0.35ex}%
    \hat{\vphantom{\rule{1pt}{\dhatheight}}%
    \smash{\hat{#1}}}}

\title{Comparison of Bayesian and particle swarm algorithms for hyperparameter optimisation in machine learning applications in high energy physics}
\author{\begin{tabular}{c@{\qquad}c}
  \authorblock{Tani, Laurits$^1$\\ \texttt{laurits.tani@cern.ch} } \and
  \authorblock{Veelken, Christian$^1$ \\ \texttt{Christian.Veelken@cern.ch}} \and 
\end{tabular}}
\date{%
    {\footnotesize $^1$\emph{National Institute Of Chemical Physics And Biophysics (NICPB), Akadeemia tee 23, 12618 Tallinn, Estonia}\\}
}

\begin{document}

    \twocolumn[
      \begin{@twocolumnfalse}
        \maketitle
        \begin{abstract}
            When using machine learning (ML) techniques, users typically need to choose a plethora of algorithm-specific parameters, referred to as hyperparameters.
            In this paper, we compare the performance of two algorithms, particle swarm optimisation (PSO) and Bayesian optimisation (BO), for the autonomous determination of these hyperparameters in applications to different ML tasks typical for the field of high energy physics (HEP). Our evaluation of the performance includes a comparison of the capability of the PSO and BO algorithms to make efficient use of the highly parallel computing resources that are characteristic of contemporary HEP experiments.
        \end{abstract} 
      \end{@twocolumnfalse}
    ]
    \maketitle


\section{Introduction}
    Machine learning (ML) methods often aid the analysis of the vast amounts of data that are produced by contemporary high energy physics (HEP) experiments.
    The ML algorithms often feature tunable parameters, referred to as hyperparameters, which need to be chosen by the user and often have a significant effect on the algorithm's performance.
    In a previous publication~\cite{tani2021evolutionary} we presented two different algorithms, particle swarm optimisation (PSO) and the genetic algorithm, for the autonomous determination of these hyperparameters.
    In the present paper we compare the performance of the PSO algorithm, the more promising of the two algorithms studied in our previous publication, to the performance of the Bayesian optimisation (BO)~\cite{BayesPaper1,BayesPaper2,BayesPaper3,BayesPaper4,BayesBook} algorithm. The latter is widely used for the task of finding optimal hyperparameter values in the context of ML applications since the pioneering work of Refs.~\cite{snoek2012practical,BayesMLPaper,thornton2013autoweka}.
    The ``asynchronous successive halving algorithm'' (ASHA)~\cite{li2020massively} is an alternative
    algorithm for optimising the values of hyperparameters, which is popular in the ML community outside the field of HEP.
    
    As in our previous publication, we formulate the task of determining the set of optimal hyperparameter values as a function maximisation problem.
    More specifically, given an ML algorithm $\mathcal{A}$, we seek to find a point $h$ in the space $\mathcal{H}$ of hyperparameters, such that the performance of the ML algorithm $\mathcal{A}$ attains its maximum at this point:
    
    \begin{equation*}
        \hat{h} = \argmax_{h \, \in \, \mathcal{H}} \, s(h) \, ,
    \end{equation*}
    
    where the objective (or ``score'') function (OF) $s(h)$ quantifies the performance of the ML algorithm $\mathcal{A}$, and the point $h$ at which $s(h)$ attains its maximum is denoted by the symbol $\hat{h}$. 
    We compare the performance of the PSO and BO algorithms on two benchmark tasks,
    the task of finding the minimum of the Rosenbrock function~\cite{shang2006note},
    and on a typical data analysis task in the field of HEP, the ``ATLAS Higgs boson machine learning challenge''~\cite{adam2015higgs}.
    
    An important aspect in applications of ML algorithms in the field of HEP is an algorithm's capability to make efficient use of massively parallel computing facilities.
    A single training of an ML algorithm on a single machine may take several hours, days, or, in extreme cases, even weeks.
    In the context of the hyperparameter optimisation task, such a single training corresponds to a single evaluation of the OF $s(h)$.
    In order for the hyperparameter optimisation task to finish within an acceptable time, different evaluations of the OF, \ie different ML trainings, need to be executed in parallel.
    The computing facilities of contemporary HEP experiments typically allow users concurrent access to hundreds, sometimes even thousands, of machines.
    It is therefore of high practical relevance whether the PSO and BO algorithms can organise the hyperparameter optimisation task such that hundreds of ML trainings can be executed in parallel.
    We find that both algorithms fulfill this requirement, but do exhibit some differences in performance compared to the case that all ML trainings are executed sequentially on a single machine.
    
    The manuscript is structured as follows: In Section~\ref{sec:bayes}, we present the main concepts of the BO algorithm. The PSO algorithm is described in Ref.~\cite{tani2021evolutionary}.
    In Section~\ref{sec:performance}, we compare the performance of the PSO and BO algorithms on the two benchmark tasks.
    The study of the parallelisation capability of the two algorithms is presented in Section~\ref{sec:parallelization}.
    We conclude the paper with a summary in Section~\ref{sec:summary}.

\section{Bayesian Optimisation}\label{sec:bayes}
    The BO algorithm is designed to facilitate the numerical maximisation of objective functions $s(h)$ which are time-consuming to evaluate and for which the analytical form is in general not known. The BO algorithm further allows to solve the maximisation task without using derivative information on $s(h)$. 
    
    This is achieved by performing the maximisation not on $s(h)$ directly, but on an approximation of $s(h)$, which is referred to as the ``surrogate'' function (SF). The SF is chosen such that it is fast to evaluate and its analytic form, including its derivative, is known. 
    We use a Gaussian process~\cite{rw2005gaussprocess} with the Mat\'ern kernel~\cite{genton2002matern} for the SF in this paper.
    For each point $h \, \in \, \mathcal{H}$, the SF provides two values: an estimate for the value $s(h)$ of the OF and a confidence interval. The latter represents an estimate of the accuracy of the approximation of the OF by the SF at this point.
    
    The numerical maximisation of the OF is performed by an iterative procedure. 
    Each iteration consists of two steps:
    The first step consists of finding the next point $h$ for which to perform the time-consuming evaluation of the OF. The task of finding this point is performed by an ``acquisition function'' (AF). The inputs to the AF consist of the estimate for the values $s(h)$ provided by the SF and the estimated accuracy of the approximation of the OF by the SF at this point.
    In the second step, the SF is updated with the information of the value of the OF at the point $h$, in order to improve the accuracy of the approximation.
    Each evaluation of the OF thus serves two purposes: 
    first, to find points $h$ where the OF attains a higher value $s(h)$ compared to previously found points and second, to improve the accuracy of the approximation of the OF by the SF.
    We use the expected improvement (EI)~\cite{jones1998ei} for the AF in this paper.
    The time-consuming evaluation of the OF is performed at the point where the AF reaches its maximum.
    The search for the maximum of the AF is performed numerically, using the Broyden–Fletcher–Goldfarb–Shanno (BFGS)~\cite{broyden1970bfgs,fletcher1970bfgs,goldfarb1970bfgs,shanno1970bfgs} algorithm.
    The numeric search for the maximum takes advantage of the fact that both the SF and the AF are fast to evaluate,
    which allows multiple evaluations of the SF and AF to be made in order to find the best point for for which to make the time-consuming next evaluation of the OF.
    The EI acquisition function has a parameter, denoted by the symbol $\xi$, that allows to regulate how much importance is given to finding points $h$ where the OF attains a high value (``exploitation'') versus finding points which improve the accuracy of the approximation (``exploration'').
    The former are typically located in the neighbourhood of previously found points, while the latter are typically located in previously unexplored regions of the space $\mathcal{H}$.
    Higher values of $\xi$ result in more exploration and lower values in more exploitation.
    The update of the SF after each evaluation of the OF is the feature to which the Bayesian optimisation algorithm owes its name.
    The SF is referred to as ``prior function'' before evaluating the OF and as ``posterior function'' afterwards.
    
    The BO algorithm is started by evaluating the OF at an initial set of points and fitting the SF to the values $s(h)$ of the OF at these points.
    This initial set of points is chosen with the objective of populating the space $\mathcal{H}$ uniformly.
    We use the Latin hypercube~\cite{mckay1979latinhcube} algorithm to obtain this initial set of points.
    After fitting the SF to the values $s(h)$ of the OF at these points, the BO algorithm enters the iterative phase:
    Given the SF, the BFGS algorithm is used to find the location where the AF attains its maximum.
    The OF is then evaluated at this point and the SF is updated.
    These steps are repeated until either the algorithm has converged, indicated by changes in the value of the OF that fall below a given threshold,
    or the computing resources are exhausted, \ie  a maximum number of iterations or a computing-time limit is reached.
    
\begin{figure}[H]
        \centering
        \includegraphics[width=0.8\columnwidth]{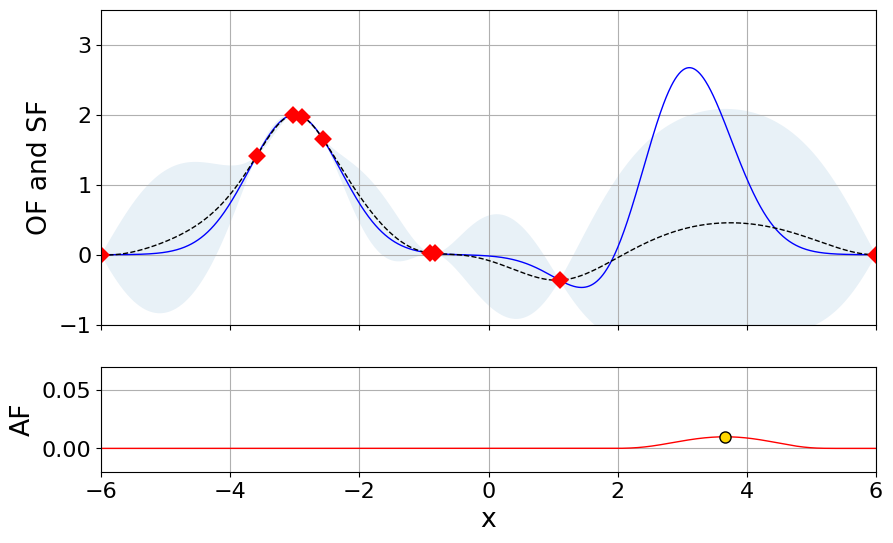}
        \caption{Example for the operation of the BO algorithm: iteration 7.}
        \label{fig:bayes3}
    \end{figure}

\begin{figure}[H]
        \centering
        \includegraphics[width=0.8\columnwidth]{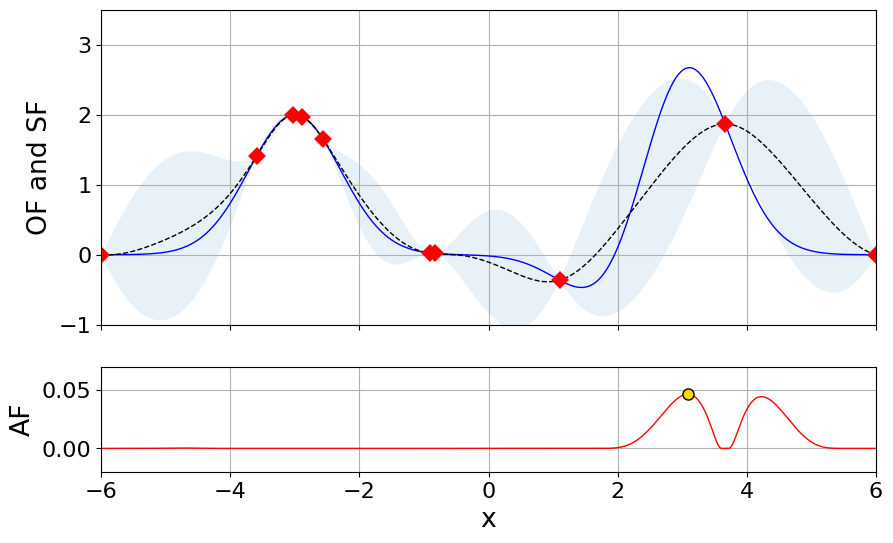}
        \caption{Example for the operation of the BO algorithm: iteration 8.}
        \label{fig:bayes4}
    \end{figure}

    The operation of the BO algorithm is illustrated by means of an example in Figs.~\ref{fig:bayes3} and~\ref{fig:bayes4}, which visualise the seventh and eighth iteration of the BO algorithm, respectively.
    The OF is represented by the solid blue line and the SF by the dashed black line in the upper part of each figure.
    The red diamond-shaped markers indicate the points where the OF has been evaluated previously (including $2$ initial points, which were chosen randomly in this example).
    The shaded blue area represents the confidence interval that quantifies the accuracy of the approximation of the OF by the SF.
    The AF is represented by the solid red line in the lower part of the figures.
    The yellow circle along the red line indicates the maximum of the AF, \ie the point $h$ where the next evaluation of the OF is performed.
    
    The BO algorithm has been implemented as described above by the authors.
    We have validated our implementation by comparing its performance to the implementation of the BO algorithm included in the \textrm{scikit-optimize}~\cite{scikit-learn} package.
    
    The BO algorithm that we described above allows to evaluate the OF at one point $h$ per iteration of the algorithm. This is well suited for executing the BO algorithm sequentially on a single machine.
    For the purpose of optimising the hyperparameters of ML algorithms the training of which is performed on massively parallel computing facilities, the iterative phase of the BO algorithm needs to be extended to provide multiple points $h$ per iteration of the BO algorithm.
    This extension of the BO algorithm is non-trivial and still a field of active research~\cite{frazier2018tutorial}.
    In this paper, we follow the approach described in Ref.~\cite{ginsbourger:hal-00260579}, referred to as ``multi-points expected improvement'' ($q$-EI), and use the implementation provided by Ref.~\cite{wang2020parallelbayes} when running the hyperparameter parameter optimisation on multiple machines in parallel.

\section{Performance}\label{sec:performance}
    In this section, the performance of the PSO and BO algorithms are compared on the two benchmark tasks: the finding of the minimum of the Rosenbrock function and the ATLAS Higgs boson machine learning challenge.
    
    \subsection{Rosenbrock function}\label{sec:rosenbrock}  
        The Rosenbrock function~\cite{shang2006note} is a widely used trial function for evaluating the performance of function minimisation algorithms.
        It is defined as:
        \begin{equation}\label{eq:rosenbrock}
            R(x, y) = (a - x)^2 + b(y - x^2)^2 \, .
        \end{equation}
        It's domain is the x--y plane and it depends on two parameters $a$ and $b$.
        The global minimum of the Rosenbrock function is located at $(x, y) = (a, a^2)$.
        We choose the two parameters to be $a = 1$ and $b = 10$, resulting in the minimum to be located at $(x, y) = (1, 1)$ and the value of the Rosenbrock function at the minimum to be $R(1, 1) = 0$.
        See Fig. 3 in Ref.~\cite{tani2021evolutionary} for a visualisation of the Rosenbrock function around the minimum.
        
        We treat the task of finding the minimum of the Rosenbrock function as a two dimensional hyperparameter optimisation problem. The optimal set of hyperparameters is scanned in the range of $[-500, +500] \times [-500, +500]$.
        
        Both the BO and the PSO algorithms were executed for $30$ iterations.
        During each iteration, $100$ different hyperparameter sets were evaluated in parallel.
        The parameter settings for both algorithms are given in Tables~\ref{tab:parameter_settings_BO} and~\ref{tab:parameter_settings_PSO}.
        The minimisation of the Rosenbrock function was repeated for $1000$ trials, using a different random number seed for each trial.
        
        \begin{table}[H]
        \centering
        \begin{tabular}{c c}
            Parameter & Value \\
            \hline
            $N^{points}_{init}$ & $100$ \\
            $\xi$ & $0.01$ \\
        \end{tabular}
        \caption{
          Parameter settings for the BO algorithm. The parameter $N^{points}_{init}$ refers to the number of points, obtained from the Latin hypercube algorithm, that are used to initialise the BO algorithm. The parameter $\xi$ regulates the relative importance of exploitation versus exploration of the EI acquisition function.
        }
        \label{tab:parameter_settings_BO}
        \end{table}

        \begin{table}[H]
        \centering
        \begin{tabular}{c c}
            Parameter & Value \\
            \hline
            \emph{$N_{info}$} & $10$ \\
            \emph{$c_1$} & $1.62$ \\
            \emph{$c_2$} & $1.62$ \\
            \emph{$w_{min}$} & $0.4$ \\
            \emph{$w_{max}$} & $0.8$ \\
        \end{tabular}
        \caption{
          Parameter settings for the PSO algorithm. The parameters are described in Ref.~\cite{tani2021evolutionary}.
        }
        \label{tab:parameter_settings_PSO}
        \end{table}

\begin{figure}[H]
            \centering
            \includegraphics[width=0.8\columnwidth]{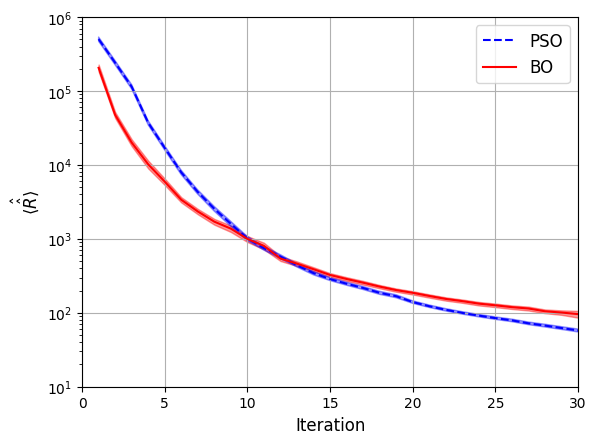}
            \caption{
              Minimum of the Rosenbrock function found by the BO and PSO algorithms, as function of the number of iterations. The lines represent the average performance over $1000$ trials and the shaded bands the variation (standard deviation) of the performance.
            } 
            \label{fig:evol_mean}
        \end{figure}

        \paragraph{Performance} We denote the location of the minimum found in each trial $i$
        by the symbol $\doublehat{h}_{i}$ and the value of the Rosenbrock function at these points by the symbol $\doublehat{R}_{i} = R(\doublehat{h}_{i})$. The average of these values over the $1000$ trials, $\langle\doublehat{R}\rangle = \frac{1}{1000} \cdot \sum_{i=1}^{1000} \doublehat{R}_{i}$, is shown in Fig.~\ref{fig:evol_mean}.
        For the first $10$ iterations, the BO algorithm converges faster to the minimum than the PSO algorithm, but improves less rapidly than the PSO algorithm for more than $10$ iterations.
        This difference in the rate of convergence is expected, since the BO algorithm has been developed for applications where the number of evaluations of the OF is in the order of a few hundred to a thousand~\cite{frazier2018tutorial} (here it is $3000$, given by the product of $30$ iterations times $100$ different hyperparameter sets that are evaluated in parallel per iteration).
        Neither the BO nor the PSO algorithm converges to the true minimum $R = 0$ within $30$ iterations. We have shown in Ref.~\cite{tani2021evolutionary} that the PSO algorithm converges to the true minimum after $10^{4}$ iterations.
        The shaded band in Fig.~\ref{fig:evol_mean} represents the standard deviation of the $\doublehat{R}_{i}$ over the $1000$ trials.
         
    \subsection{The ATLAS Higgs Boson machine learning challenge}\label{sec:HBC}
        The \emph{ATLAS Higgs Boson machine learning challenge (HBC)}~\cite{adam2015higgs} constitutes the second benchmark task for evaluating the performance of the BO and PSO algorithms.
        As in our previous publication~\cite{tani2021evolutionary}, we choose a boosted decision tree (BDT), implemented in the \textrm{XGBoost}~\cite{chen2016xgboost} package, for the ML algorithm. The hyperparameter optimisation is performed with respect to the $7$ hyperparameters given in Table~3 of Ref.~\cite{tani2021evolutionary}. During the optimisation, each of the $7$ hyperparameters is restricted to be within the range given in Table~\ref{tab:hbc_parameter_ranges}.
        
        \begin{table}[H]
            \centering
            \begin{tabular}{c c c}
                Hyperparameter & min & max \\
                \hline
                \emph{num-boost-round} & $1$ & $500$ \\
                \emph{learning-rate} & $10^{-5}$ & $1$ \\
                \emph{max-depth} & $1$ & $6$ \\
                \emph{gamma} & $0$ & $5$ \\
                \emph{min-child-weight} & $0$ & $500$ \\
                \emph{subsample} & $0.8$ & $1$ \\
                \emph{colsample-bytree} & $0.3$ & $1$ \\
            \end{tabular}
            \caption{
              Minimum and maximum values of the hyperparameters for the HBC. 
              The hyperparameters are detailed in Ref.~\cite{adam2015higgs}.
            }
            \label{tab:hbc_parameter_ranges}
        \end{table}
        
        The optimisation of the hyperparameters by the BO and PSO algorithms was run for $30$ iterations. $70$ BDTs, initialised with different random number seeds, were trained in parallel per iteration. The parameter settings used for the BO and PSO algorithm are the same as for the task of finding the minimum of the Rosenbrock function and are given in Tables~\ref{tab:parameter_settings_BO} and~\ref{tab:parameter_settings_PSO}, except for the parameter $N_{info}$ of the PSO algorithm, which was set to $7$ for the HBC task.
        The $550000$ signal and background events provided by the organisers of the HBC are split into training and test sets as described in Ref.~\cite{tani2021evolutionary}. 
        The ``modified approximate mean significance'' (d-AMS) defined by Eq.~(8) of Ref.~\cite{tani2021evolutionary} was used as objective function for the BDT training.
        The d-AMS scores were computed setting the coefficient $\kappa$ that controls the penalty term against overtraining to $0.3$.
        The hyperparameter optimisation was repeated for $100$  trials.
        
        \subsubsection{Results}
        The average hyperparameter values found by the BO and PSO algorithms over the $100$ trials and the standard deviation of these values are reported in Table~\ref{tab:bo_pso_repetition}.
        The values found by both algorithms are similar.
        The \emph{gamma} hyperparameter has little effect on the d-AMS score and is thus not well constrained by the optimisation.
        
        \begin{table}[H]
            \centering
            \begin{tabular}{l cc cc}
            \multirow{2}{3.1cm}{Hyperparameter} & \multicolumn{2}{c}{BO} & \multicolumn{2}{c}{PSO} \\
            & $\mu$ & $\sigma$ & $\mu$ & $\sigma$ \\
            \hline
            \emph{num-boost-round} & $364.3$ & $85.2$ & $413.7$ & $85.4$ \\
            \emph{learning-rate} & $0.126$ & $0.041$ & $0.089$ & $0.024$ \\
            \emph{max-depth} & $3.9$ & $0.7$ & $4.5$ & $0.7$ \\
            \emph{gamma} & $1.78$ & $2.00$ & $2.81$ & $1.47$ \\
            \emph{min-child-weight} & $352.7$ & $130.1$ & $440.4$ & $72.4$ \\
            \emph{subsample} & $0.861$ & $0.081$ & $0.871$ & $0.066$ \\
            \emph{colsample-bytree} & $0.852$ & $0.268$ & $0.859$ & $0.145$ \\
            \end{tabular}
            \caption{
              Average values ($\mu$) of the hyperparameters found by the BO and PSO algorithms and their standard deviation ($\sigma$) for the HBC task.
            }
            \label{tab:bo_pso_repetition}
            \end{table}

\begin{figure}[H]
                \centering
                \includegraphics[width=0.8\columnwidth]{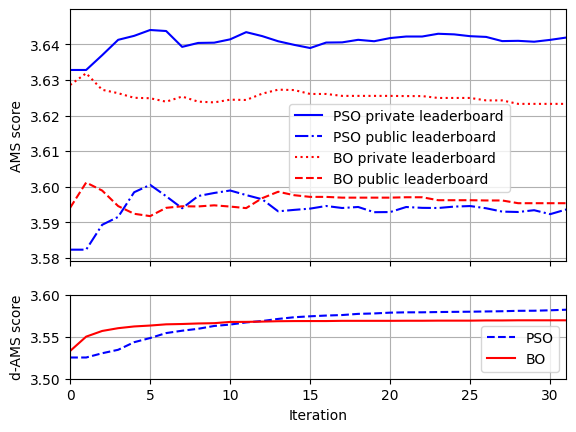}
                \caption{
                  Evolution of the AMS score as function of the number of iterations for the HBC task. The lines represent the average performance over $100$ trials.
                }
                \label{fig:local_HBC_evol}
            \end{figure}

            The performance of BDTs is evaluated on two separate samples of events, referred to as the \emph{public} and \emph{private} leaderboard samples~\cite{tani2021evolutionary,adam2015higgs}. 
            Following Ref.~\cite{tani2021evolutionary}, the performance is quantified by the ``approximate mean significance'' (AMS) score. The latter is averaged over the $100$ trials.
            The motivation for using the d-AMS score as objective function for the training, while the (final) performance is quantified using the AMS score is detailed in Ref.~\cite{tani2021evolutionary}.
            The evolution of the average performance as function of the number of iterations is shown in Fig.~\ref{fig:local_HBC_evol}.
            The bottom part of the figure shows the evolution of the d-AMS score, the objective function used for the BDT training. 
            While the d-AMS score keeps increasing monotonously with more iterations, the AMS scores on the public and private leaderboard samples reach a plateau already after $5$--$10$ iterations.
            In subsequent iterations, the AMS scores start to fluctuate around the plateau. The magnitude of the fluctuations is of $\mathcal{O}(10^{-2})$.
            The differences in performance between the public and private leaderboard samples is compatible with a statistical fluctuation of the signal and background events contained in these samples~\cite{adam2015higgs}.
            The BO and PSO algorithms achieve a similar performance on the HBC task.

\section{Parallelisation capability}\label{sec:parallelization}
    The capability of the BO and PSO algorithms to make efficient use of parallel computing resources is studied on the task of finding the minimum of the Rosenbrock function.
    We denote by $N_{parallel}$ the number of points $h$ in hyperparameter space that are evaluated in parallel per iteration of the BO and PSO algorithms.
    The value of $N_{parallel}$ represents the number of ML trainings that are executed in parallel on different machines.
    Ideally, the duration (``wall time'') of the hyperparameter optimisation task should decrease inversely proportional to $N_{parallel}$, except for some (small) overhead imposed by the BO and PSO algorithms.
    
    \paragraph{Bayesian optimisation} Based on the literature \cite{frazier2018tutorial}, one expects the BO algorithm to perform best when all ML trainings are performed sequentially on a single machine, with one point $h$ in hyperparameter space evaluated per iteration.
    We have compared the performance, quantified by the values of $\langle\doublehat{R}\rangle$, obtained when executing our implementation of the BO algorithm (described in Section~\ref{sec:bayes}) sequentially on a single machine ($N_{parallel} = 1$) with the performance obtained by the $q$-EI algorithm from Ref.~\cite{wang2020parallelbayes},
    with the latter running either $N_{parallel} = 2$, $5$, $10$, $25$, $50$, $100$, $250$, $500$ or $1000$ ML trainings in parallel.
    We find that the values of $\langle\doublehat{R}\rangle$ are very similar in all cases and mainly depend on the total number of evaluations of the Rosenbrock function, given by the product of the number of iterations times $N_{parallel}$.
    For $3000$ evaluations, the value of $\langle\doublehat{R}\rangle$ amounts to about $10^{2}$ for all values of $N_{parallel}$ that we have tried.
    
    The computing (CPU) time overhead imposed by the BO algorithm for updating the SF and finding the maximum of the AF amounts to about $50$ CPU hours on a $2.30$~GHz Intel\TReg~Xeon\TReg~E$5$-$2695$ v$3$ processor. The overhead does not vary much as function of $N_{parallel}$. The task of updating the SF and finding the maximum of the AF needs to run (as a ``supervisor'' task) on a single machine. The overhead is sizeable compared to the computing time spent on performing $3000$ evaluations of the Rosenbrock function, which only takes $0.06$ CPU seconds, but becomes less important for ``real'' ML training applications: The total computing time spent on the HPC task (with $30$ iterations and $70$ ML trainings running in parallel) amounts to about $500$ CPU hours.
    
    \paragraph{Particle swarm optimisation} In contrast to the BO algorithm, the capability to make efficient use of parallel computing resources is an intrinsic feature of the PSO algorithm.
    We find that the performance of the PSO algorithm depends on the value of $N_{parallel}$, which corresponds to the number of particles in the swarm, in a non-trivial manner.
    Both too small and too large values of $N_{parallel}$ degrade the performance of the PSO algorithm compared to the optimal value.
    We find that the optimal value of $N_{parallel}$ amounts to about $2\%$ times the total number ($N_{tot}$) of evaluations of the Rosenbrock function, \ie the best performance of the PSO algorithm is achieved when using $N_{parallel} = 0.02 \cdot N_{tot}$ particles in the swarm and a fixed number of $50$ iterations (up to $10000$ evaluations of the Rosenbrock function that we have tried).
    We recommend to set the parameter $N_{info}$ of the PSO algorithm to $10\%$ times $N_{parallel}$.
    
    We find that the CPU time overhead imposed by the PSO algorithm (spent on updating the positions and momenta of the particles in the swarm) is negligible for both benchmark tasks.
    
\begin{figure}[H]
        \centering
        \includegraphics[width=0.8\columnwidth]{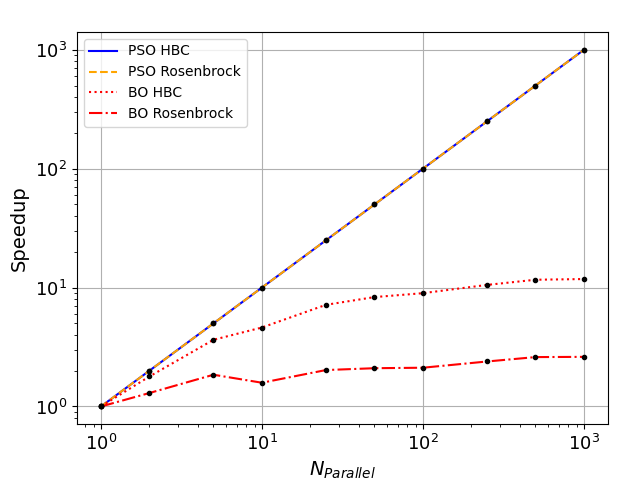}
        \caption{Amdahl's law: Parallelisation properties of the BO and PSO algorithms on the two benchmark tasks. The two curves for the PSO algorithm are very close.}
        \label{fig:amdahl}
    \end{figure}

    The CPU time overhead limits the ``speedup'' (reduction in wall time) that one can achieve by increasing the number of points in hyperparameter space that are evaluated in parallel. The effect is referred to as ``Amdahl's law''~\cite{amdahl} in the literature and visualized in Fig.~\ref{fig:amdahl}. An ideal algorithm with negligible overhead would achieve a speedup factor equal to $N_{parallel}$ in this figure. The speedup achieved by the PSO algorithm on both benchmark tasks is very close to the ideal case. The CPU time overhead limits the speedup factor achievable by the BO algorithm to about $3$ for the task of finding of the minimum of the Rosenbrock function and to about $12$ for the HBC task. 

\section{Summary}\label{sec:summary}
    We have compared the performance of two autonomous algorithms for the optimisation of hyperparameters,
    Bayesian optimisation (BO) and particle swarm optimisation (PSO), on two benchmark tasks typical for ML applications in the field of high energy physics: the task of finding the minimum of the Rosenbrock function and the ATLAS Higgs boson machine learning challenge.
    
    We find that the BO algorithm performs better than the PSO algorithm when the total number of evaluations of the Rosenbrock function (equivalent to the number of ML trainings) is of the order of a few hundred to a few thousand. If the number of evaluations (ML trainings) is large, the PSO algorithm outperforms the BO algorithm.
    
    The capability of both algorithms to make efficient use of parallel computing resources is good. In particular, we find that the ``multi-points expected improvement'' of the BO algorithm provides similar performance when running on parallel computing resources compared to executing the BO algorithm sequentially on a single machine.
    In the case of the PSO algorithm, we found that the best performance is achieved by setting the number of particles in the swarm to $2\%$ times the total number of function evaluations (ML trainings) and using a fixed number of $50$ iterations.
    
    We found that the BO algorithm may add a significant computational overhead to the task of finding the optimal hyperparameter values, while for the PSO algorithm the overhead is insignificant.

\section*{Acknowledgements}
    We would like to thank our colleague Joosep Pata for helpful discussions.
    This work has been supported by the Estonian Research Council grant PRG445.

\section*{Data availability}
Data used for the Higgs Boson Machine Learning Challenge is available made available by the organizers of that competition at \url{http://opendata.cern.ch/record/328}. The competition itself can be found at \url{https://www.kaggle.com/c/higgs-boson}. Software used to produce the results can be found in \cite{laurits_tani_2023_8171318}.

\section*{Author contributions (CRediT)}
\textbf{Laurits Tani}: Conceptualization, Data curation, Formal analysis, Investigation, Methodology, Software, Validation, Visualization, Writing – original draft, Writing – review \& editing.
\textbf{Christian Veelken}: Conceptualization, Funding acquisition, Supervision,  Writing – original draft, Writing – review \& editing.

\bibliography{main}

\end{document}